# Multi-axis Accelerometry and Rotation Sensing using a Point Source Atom Interferometer

Jinyang Li[1], Gregório R. M. da Silva[1], Jason Bonacum[2,3], Timothy Kovachy[1], and Selim M. Shahriar[1,2]

[1] Department of Physics and Astronomy, Northwestern University, Evanston, IL 60208, USA
[2] Department of Electrical and Computer Engineering, Northwestern University, Evanston, IL 60208, USA
[3] Digital Optics Technologies, Rolling Meadows, IL 60008, USA

## Abstract

A point source atom interferometer (PSI) is a device where atoms are split and recombined by applying a temporal sequence of Raman pulses during the expansion of a cloud of cold atoms behaving approximately as a point source. Unlike a conventional light pulse atom interferometer, the PSI can produce a signal that corresponds to multi-axis rotation only, independent of acceleration. In addition, it can be used to measure acceleration along one direction, independent of rotation. Here, we describe a modified PSI that can be used to measure multi-axis rotation and multi-axis acceleration. Specifically, this type of PSI can be used to measure two-axes rotation around the directions perpendicular to the light pulses, as well as the acceleration in all three directions, with only one pair of Raman beams. Using two pairs of Raman beams in orthogonal directions sequentially, such a scheme would enable the realization of a complete atom interferometric inertial measurement unit.

# 1. Introduction

Atom interferometry offers the potential to deliver a high-performance, compact, and robust inertial measurement unit that is suitable for inertial navigation applications. Critical requirements for such an atomic-interferometric inertial measurement unit (AIMU) include high sensitivity to rotations and accelerations, along all three directions, as well as the ability to distinguish between signals arising from rotations and accelerations. Existing techniques for atomic interferometry [1] can, in principle, be used to realize such an AIMU. However, it is likely to be very large, since several independent devices may be needed for measuring rates of rotation and acceleration in all three directions. Here, we propose a new technique, employing a point source atom interferometer (PSI) [2,3,4,5,6], that may enable the realization of an AIMU using a single device, thus making it potentially very compact.

A point source atom interferometer (PSI) is a device where atoms are split and recombined by applying a temporal sequence of laser pulses during the expansion of a cloud of cold atoms behaving approximately as a point source. The pulses consist of a pair of counter-propagating laser beams that drive two-photon Raman transitions [7], as shown in Figure 1(c), serving as the beam splitters and mirrors for a Mach–Zehnder light-pulse atom interferometer [8,9,10,11,12,13,14], as shown in Figure 1(a). Both the rotation perpendicular to the light pulses and the acceleration parallel to the light pulses will introduce a phase shift in the PSI. The interferometer phase response to rotation scales linearly with the initial velocity transverse to the axis of the Raman beams, while the response to acceleration is independent of the atomic velocity. Because of this difference, the PSI allows the signals due to the rotation and the acceleration to be distinguished. The PSI can also determine both components of the rotation vector that are orthogonal to the laser pulses, thus realizing a two-axis gyroscope and one-axis accelerometer. It should be noted that there are other techniques that can also distinguish between rotation and acceleration [10,11,12,13,14]. However, a key practical advantage of the PSI is that it only requires a single atom cloud and Raman beams along a single axis, in contrast to other methods.

The conventional PSI employs three sequential Raman pulses, which produces a non-zero Sagnac area. It is also possible to add another pulse in the sequence, thus producing a zero Sagnac area in the absence of any acceleration, as shown in Figure 1(b). Here, we show that by introducing a bias rotation, employing a rotational retro-reflection mirror, as shown in Figure 1(d), it is possible to measure the acceleration along each of the two axes perpendicular to the light pulses with the four-pulse interferometer. We note that this method of introducing an artificial rotation is also widely used in conventional PSI experiments [2,3,4,5]. Thus, by alternating between the three-pulse and four-pulse sequences for the PSI, along with the application of the bias rotation, it is possible to measure the acceleration in all three directions and the rotations in the two directions perpendicular to the light pulses. Of course, it is also possible to apply the Raman pulses along two different, orthogonal directions alternatingly, thereby realizing two different sets of three-pulse and four-pulse PSIs. Such a system would enable the measurement of rotations and accelerations along all three axes, thereby realizing a three-axis AIMU.

In a recent paper [6], we have pointed out that the sensitivity of a PSI can be increased significantly, by a factor as large as 40 for experimentally accessible parameters, using the process of large momentum transfer. Both three-pulse and four-pulse versions of the PSI can be augmented by this process, thereby enhancing the sensitivity of both rotation sensing and accelerometry significantly.

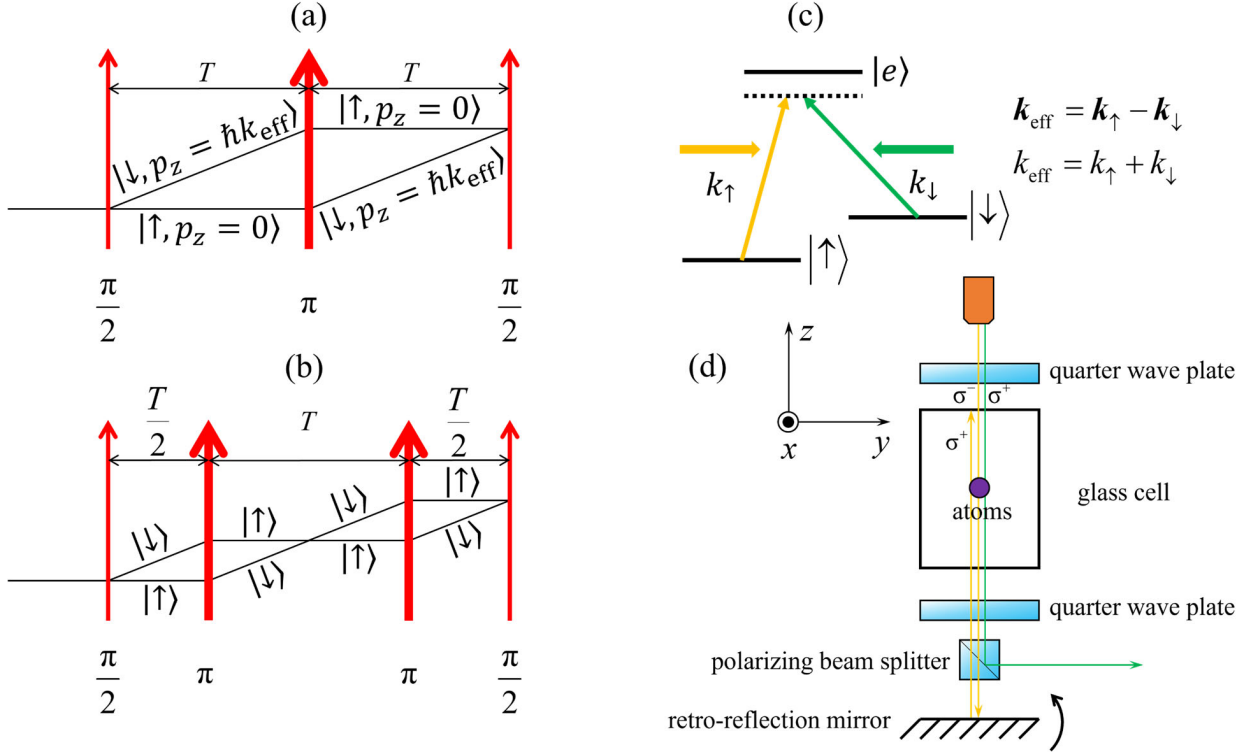

Figure 1. (a) Three-pulse Mach–Zehnder atom interferometer. (b) Four-pulse atom interferometer. (c) Schematic diagram of the atom model and the pair of counter-propagating Raman beams. (d) Schematic diagram of the PSI apparatus. The two perpendicularly linearly polarized Raman beams go downwards from the same fiber port. The upper quarter waveplate turn their polarizations into right-circular ($\sigma^+$) and left-circular ($\sigma^-$) ones, respectively. After they pass through the glass cell, another quarter waveplate turn them back to perpendicularly linearly polarized light. One beam passes through the polarizing beam slitter (PBS) while the other is reflected by the PBS. The beam passing through the PBS is retro-reflected by the mirror mounted on a rotation stage. Although both the counter-propagating and co-propagating Raman beams exist, we can excite only the counter-propagating transitions by shifting the frequency difference between the two beams, since the counter-propagating transitions are far more velocity-sensitive than the co-propagating transitions.

The rest of the paper is organized as follows. In Section 2, we analyze the signal resulting from a three-pulse PSI, and show explicitly how it can be used to reveal the acceleration along the direction of the light-pulses, in addition to the rates of rotation along the two axes perpendicular to this direction. In Section 3, we analyze the signal for the four-pulse configuration of the PSI, and show how the application of a bias rotation enables the measurement of accelerations perpendicular to the direction of the light-pulses. In Section 4, we discuss how the sensitivity of both rotation sensing and accelerometry employing this approach can be enhanced by using the technique of large momentum transfer. We summarize the findings and implications thereof in Section 5. Technical details of how to measure acceleration using the three-pulse PSI are presented in the Appendix.

## 2. Signal of the three-pulse PSI

To discuss the signal of the three-pulse PSI, we first introduce the model of an atom necessary for describing the Raman transition. Each atom is modeled as a three-level system with the two ground states denoted as $|\uparrow\rangle$ and $|\downarrow\rangle$, and the excited state denoted as $|e\rangle$. In practice, the two ground states are normally the $m_F = 0$ Zeeman substates of the two hyperfine ground states of an alkali atom. The Raman beam coupling the two states $|\uparrow(\downarrow)\rangle$ and $|e\rangle$ is denoted as $\boldsymbol{k}_{\uparrow(\downarrow)}$, as shown in Figure 1(c). The effective wavenumber of the pair of counter-propagating Raman beams can be expressed as $\boldsymbol{k}_{\text{eff}} = (\boldsymbol{k}_\uparrow - \boldsymbol{k}_\downarrow)$ if the direction of $\boldsymbol{k}_{\text{eff}}$ is defined to be that of $\boldsymbol{k}_\uparrow$. The absolute value of the effective wavenumber can be expressed as $k_{\text{eff}} = (k_\uparrow + k_\downarrow) \approx 2k_{\uparrow(\downarrow)}$, where $k_{\uparrow(\downarrow)}$ is the absolute value of the corresponding wavenumber. The acceleration induced phase shift of the three-pulse interferometer can be expressed as $\phi_a = \boldsymbol{k}_{\text{eff}} \cdot \boldsymbol{a} T^2$, and the rotation induced phase shift as $\phi_\Omega = (\boldsymbol{k}_{\text{eff}} \times \boldsymbol{\Omega} T) \cdot \boldsymbol{r}$, where $\boldsymbol{a}$ is the acceleration, $\boldsymbol{\Omega}$ is the rotational velocity, and $\boldsymbol{r}$ is the displacement of the atoms between the first and the last π/2-pulse. The total phase shift is $\phi = (\phi_a + \phi_\Omega)$. The signal, which is the spatial distribution of the population of state $|\uparrow\rangle$, can be expressed as $(1+\cos\phi)/2$. The atoms with different velocities will have different displacements, and thus different rotation induced phase shifts. In a PSI starting with an ideal point source, the displacement of an atom is just its final position. Consequently, the signal of the PSI will be spatial fringes with a wavenumber of $k_\Omega \equiv k_{\text{eff}} \times \Omega T$. The absolute value of this wavenumber is proportional to the component of the angular velocity perpendicular to $k_{\text{eff}}$ and the orientation of the fringes is parallel to that component, as shown in Figure 2. At the position of the point source where $r = 0$, the phase shift will be $\phi_a$, which depends only on the acceleration. In summary, the wavenumber (which determines the periodicity) of the fringes indicates the angular velocity of the rotation, and the phase of the fringes at $r = 0$ indicates the acceleration, as shown in Figure 3(a1)-(a3). From another perspective, it can be stated that the acceleration causes a displacement of the fringes [15], as shown in Figure 2. To illustrate this point explicitly, consider, for example, the case when $k_{\text{eff}}$ is in the z-direction and $\Omega$ is in the x-direction, as shown in Figure 1(d). Then the phase shift can be expressed as

$$\phi = k_{\text{eff}} a_z T^2 + k_{\text{eff}} \Omega_x T y = k_{\text{eff}} \Omega_x T \left( y + \frac{a_z T}{2\Omega_x} \right) \tag{1}$$

which is shifted to the $(-y)$-direction by $a_z T / \Omega_x$.

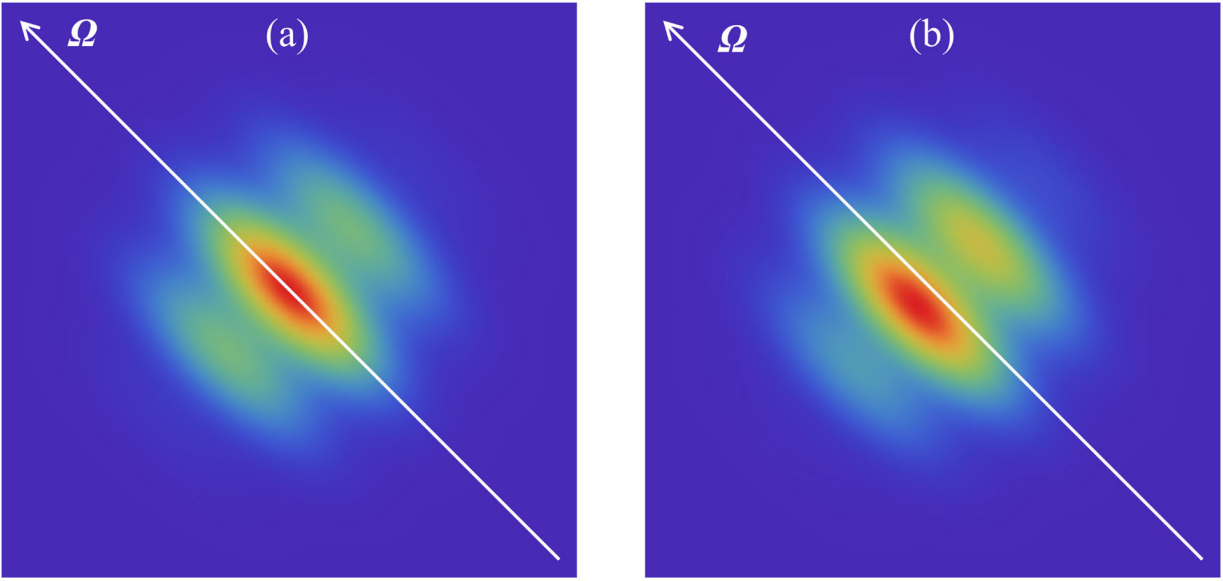

Figure 2. Example of theoretical signal of the PSI (a) with no acceleration induced phase shift and (b) with an acceleration induced phase shift of π/4. The wavenumber of the fringes is $k_{\mathrm{eff}} \times \boldsymbol{\Omega} T$. The absolute value of the wavenumber is proportional to that of the angular velocity and the orientation of the wavenumber is parallel to that of the angular velocity. Therefore, these fringes enable a two-axis measurement of rotation with the three-pulse PSI. The phase of the fringes at $r=0$ indicates the acceleration, or equivalently, the displacement of the fringes indicates the acceleration.

For concreteness of discussion, we assume that a one-to-one imaging system is used to detect, with a focal plane array (FPA), the variation of the populations in one of the two ground states. The signal can thus be expressed as $\left(2+e^{i(k_\Omega y+\phi_a)}+e^{-i(k_\Omega y+\phi_a)}\right)/4$, where $k_\Omega = k_{\mathrm{eff}}\Omega_x T$ and $\phi_a = k_{\mathrm{eff}} a_z T^2$. A fast way to extract the signals of both the acceleration and the rotation is to calculate the Fourier transform of the signal. This calculation can be implemented, for example, by a field programmable gate array (FPGA). The amplitude spectrum of the Fourier transform consists of three peaks at $k_y = 0, \pm k_\Omega$ as shown in Figure 3(b1)-(b3), and the phases of the Fourier transform at two side peaks are $\pm \phi_a$, as shown in Figure 3(c1)-(c3). Accordingly, the distance between a side peak and the central peak indicates the angular velocity and the phase of the side peak indicates the acceleration. Experimentally, extracting the information about the rotation is straightforward, while measuring the acceleration involves two complexities: there will be no side peaks in the Fourier transform in the absence of rotation and the phase of the Fourier transform depends on the choice of the origin of the coordinates. It should be noted that if the four-pulse PSI is also implemented with Raman beams propagating along the *x*-axis, for example, as described in the next section, it will yield values of the accelerations along the *y*-axis and the *z*-axis. Thus, there is no need to measure the acceleration along the *z*-axis using the three-pulse PSI. Nonetheless, we show in the Appendix how to measure the acceleration along the *z*-axis using the three-pulse PSI with the Raman beams propagating along the *z*-direction.

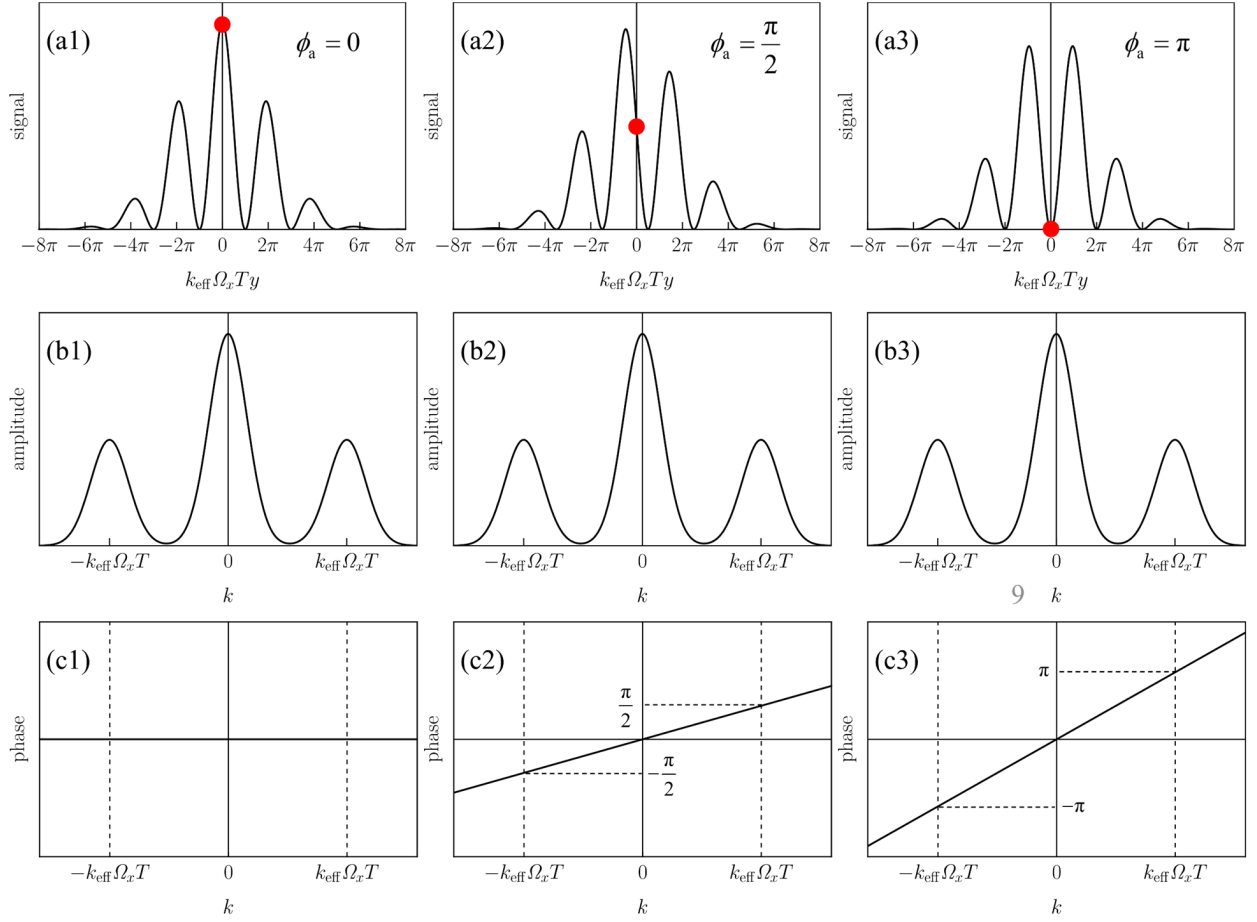

Figure 3. (a1)-(a3): Signal of the PSI in the spatial domain with different values of the acceleration induced phase shift, assuming that $\boldsymbol{k}_{\text{eff}}$ is in the $z$-direction and $\boldsymbol{\Omega}$ is in the $x$-direction. The phase of the fringes at $\boldsymbol{r} = 0$ (red dot) indicates the acceleration. (b1)-(b3): The amplitude spectrum of the Fourier transform of the signal shown in (a1)-(a3) respectively. They are all the same because the acceleration does not affect the amplitude spectrum. The distance between the central peak and a side peak is proportional to the angular velocity. (c1)-(c3): Phase spectrum of the Fourier transform of the signal shown in (a1)-(a3) respectively. The phases at the positions of the two side peaks in the amplitude spectrum indicates the acceleration phase shift.

We next analyze the sensitivity of the rotation measurement with the three-pulse PSI. We do not include the corresponding analysis for the acceleration measurement because it is not necessary to use the three-pulse PSI to measure acceleration, as discussed above. The sensitivity of the rotation measurement depends on how precisely we can determine the position of the side peak in the Fourier transform. This precision is given by the signal to noise ratio, which is $\sim \sqrt{N}$ for an ideal case, divided by the width of the side peak, which is $\sim \sigma_f^{-1}$, the reciprocal of the final size of the atomic cloud. Therefore, it can be expressed as [6] $\Delta k_\Omega^{-1} = \sqrt{N} \sigma_f$. Recalling that $k_\Omega = k_{eff} \Omega T$, we can express the sensitivity of the rotation measurement as $\Delta \Omega^{-1} = k_{eff} T \sqrt{N} \sigma_f$. For $^{85}$Rb, the effective wavenumber of the counter-propagating Raman beams is $k_{eff} = 2\pi \times 2.56 \times 10^4$ cm$^{-1}$. For a set of potentially practical parameters such as $N = 10^8$ per second, $\sigma_f = 1$ cm, and $T = 0.1$ s, the uncertainty of the rotation measurement will be $\sim 6.2$ nrad·s$^{-1}/\sqrt{\text{Hz}}$. To check the validity of this approach for estimating the sensitivity of rotation measurement, we calculate the sensitivity of an equivalent conventional atom interferometer in which all the atoms move a distance $x_0$, which equals to value of $\sigma_f$ for the PSI, in the direction perpendicular to the light pulses, during the interval between the first and the last

π/2 pulses. The sensitivity of the measurement of the rotation-induced phase shift is $\Delta\phi_\Omega^{-1} = \sqrt{N}$. With the expression of the rotation-induced phase shift $\phi_\Omega = k_{\text{eff}}\Omega T x_0$, the sensitivity of the measurement is derived to be $\Delta\Omega^{-1} = k_{\text{eff}}T\sqrt{N}x_0$, which is exactly the same as the sensitivity calculated above for $x_0 = \sigma_f$.

When augmented with the large momentum transfer process (as discussed in Section 4), which may enhance the sensitivity by a factor of as much as 40 [6], the sensitivity can potentially reach a value of $\sim 0.16 \text{ nrad}\cdot\text{s}^{-1}/\sqrt{\text{Hz}}$. This would be nearly a factor of four more sensitive than the best atom interferometric gyroscope [16], which is large and based on atomic beams, demonstrated to date.

## 3. Four-pulse interferometer with a bias rotation

With the apparatus of the PSI, we can also implement the four-pulse interferometer pulse sequence shown in Figure 1(b). The four-pulse interferometer is equivalent to two ordinary three-pulse interferometers with opposite orientations [17]. Consequently, the acceleration phase shifts of these two interferometers will cancel out in all circumstances and the Sagnac phase shifts will also cancel out in the absence of acceleration perpendicular to the Raman beams. In the presence of acceleration perpendicular to the Raman beams, the residual Sagnac phase shift will be $\phi = \left[\mathbf{k}_{\text{eff}} \times \mathbf{\Omega}(T/2)\right]\cdot(\mathbf{r}_1 - \mathbf{r}_2)$, where $\mathbf{r}_{1(2)}$ is the displacement of the atoms in the first (second) three-pulse interferometer. Substituting the displacement difference $(\mathbf{r}_2 - \mathbf{r}_1) = \mathbf{a}T^2$ into the expression of the phase shift of the four-pulse atom interferometer, we have [18,19,20] $\phi = \mathbf{k}_{\text{eff}} \cdot (\mathbf{a}\times\mathbf{\Omega})T^3/2$. This phase shift does not depend on the displacement of the atoms.

Therefore, there will be no spatially varying distribution of the atoms in state $|\uparrow\rangle$. The signal for this four-pulse interferometer is the total number of atoms in state $|\uparrow\rangle$. In the absence of the environmental rotation, a bias rotation in the $x(y)$-direction will produce a phase shift that is proportional to $a_{y(x)}$, meaning that we can measure the acceleration perpendicular to $\boldsymbol{k}_{\text{eff}}$. In the presence of an environmental rotation, we can measure the acceleration perpendicular to $\boldsymbol{k}_{\text{eff}}$ with a differential measurement. Including effects of both the environmental rotation ($\boldsymbol{\Omega}_{\text{e}}$) and the bias rotation ($\boldsymbol{\Omega}_{\text{b}}$), the phase shift can be expressed as $\phi_+ = \boldsymbol{k}_{\text{eff}} \cdot \left[\boldsymbol{a} \times (\boldsymbol{\Omega}_{\text{e}} + \boldsymbol{\Omega}_{\text{b}})\right] T^3/2$. When the bias rotation is reversed in sign, the phase shift can be expressed as $\phi_- = \boldsymbol{k}_{\text{eff}} \cdot \left[\boldsymbol{a} \times (\boldsymbol{\Omega}_{\text{e}} - \boldsymbol{\Omega}_{\text{b}})\right] T^3/2$. The relative phase shift between these two measurements is $(\phi_+ - \phi_-) = \boldsymbol{k}_{\text{eff}} \cdot (\boldsymbol{a} \times \boldsymbol{\Omega}_{\text{b}}) T^3$, a quantity that depends only on the acceleration and the bias rotation. Again, for concreteness, we assume that $\boldsymbol{k}_{\text{eff}}$ is in the z-direction. By introducing a bias rotation in the $x(y)$-direction, we can measure $a_{y(x)}$. Repeating the measurements with bias rotations in both $x$- and $y$-direction, we can measure both components of the acceleration perpendicular to $\boldsymbol{k}_{\text{eff}}$.

The sensitivity of the measurement of the phase shift is [21] $\Delta\phi^{-1} = \sqrt{N}$. Therefore, for the four-pulse interferometer, the sensitivity of the acceleration measurement is $\Delta a^{-1} = k_{\text{eff}} \Omega_{\text{b}} T^3 \sqrt{N}$. Practically, the value of $\Omega_{\text{b}} T$ should be limited to the order of 0.1. For a fixed value of $\Omega_{\text{b}} T$, the sensitivity is higher for longer $T$. Therefore, the bias rotation should be as slow as possible. For the set of practical parameters such as $N = 10^8$ per second, $\Omega_{\text{b}} T = 0.1$ and $T = 0.1$ s, employing $^{85}$Rb atoms, the uncertainty of the acceleration measurement will be ~ 63.2 nano-$g$ /√Hz, where $g$ (=9.8 m/s²) is the mean terrestrial gravitational acceleration. When augmented with the large momentum transfer process (as discussed in Section 4), which may enhance the sensitivity by a factor of as much as 40 [6], the sensitivity can potentially reach a value of ~ 1.6 nano-$g$/√Hz. This would be nearly a factor of 2.7 more sensitive than the best atom interferometric accelerometer [22], which is large and based on atomic beams, demonstrated to date.

## 4. Enhancement of Sensitivity of Rotation Sensing and Accelerometry using Large Momentum Transfer

The sensitivity of measuring both rotation and acceleration using both the three-pulse and the four-pulse PSI can be enhanced by making use of the technique of large momentum transfer [6,23,24,25,26,27]. Large momentum transfer can be realized with sequential multi-photon Bragg transitions [23,24,25] and sequential Raman transitions [6,26,27]. The technique of sequential multi-photon Bragg transitions generally produces a larger momentum transfer while having the disadvantage of requiring extremely cold atoms. In contrast, the technique of sequential Raman transitions produces a smaller momentum transfer and is subject to light shift while having the advantage of tolerating hotter atoms (atoms cooled only with polarization-gradient cooling). For the Mach–Zehnder atom interferometer, the phase shifts caused by both the acceleration and the rotation are magnified by a factor given approximately by the maximum momentum transfer divided by $\hbar k_{\text{eff}}$, in the limit where the additional time needed for applying the auxiliary pulses used for large momentum transfer is small compared to the time difference between the first and the last pulse. In reference [6], we have shown that, for the three-pulse PSI, the sensitivity of rotation sensing can be enhanced by a factor as large as 40 using large momentum transfer, for experimentally accessible parameters. It is thus expected that the sensitivity of measuring acceleration, for both the three-pulse PSI, as well as the four-pulse PSI, can also be increased by nearly the same factor using the technique of large momentum transfer.

To realize the LMT with sequential Raman transitions, the directions of the counter-propagating Raman beams are reversed repeatedly. Based on the design that the Raman beam with a particular polarization will be retro reflected while the Raman beam with the orthogonal polarization is directed away, as shown in Figure 1, the directions of the Raman beams can be reversed by simply exchanging their polarizations. Considering the practical requirement that the polarization exchange must be completed within about a millisecond, Pockels cells can be used for this purpose [26,27], as shown in Figure 4.

## 5. Discussion and Conclusion

In this paper, we propose a way of using the apparatus of the PSI to measure the acceleration in all three directions and the rotation around the two directions perpendicular to the light pulses. There are two important aspects in this proposal. The first aspect is the introduction of a bias rotation using the rotational retro-reflection mirror, which is already widely used in the PSI experiments. The second aspect is an additional pulse in the sequence, in order to realize a four-pulse PSI, which has a zero Sagnac-area in the absence of any acceleration. By applying the three-pulse interferometer pulse sequence, we can measure the acceleration parallel to the light pulses and the rotation perpendicular to the light pulses simultaneously. By applying the four-pulse interferometer and introducing a bias rotation, we can measure the acceleration in the two directions perpendicular to the light pulses individually.

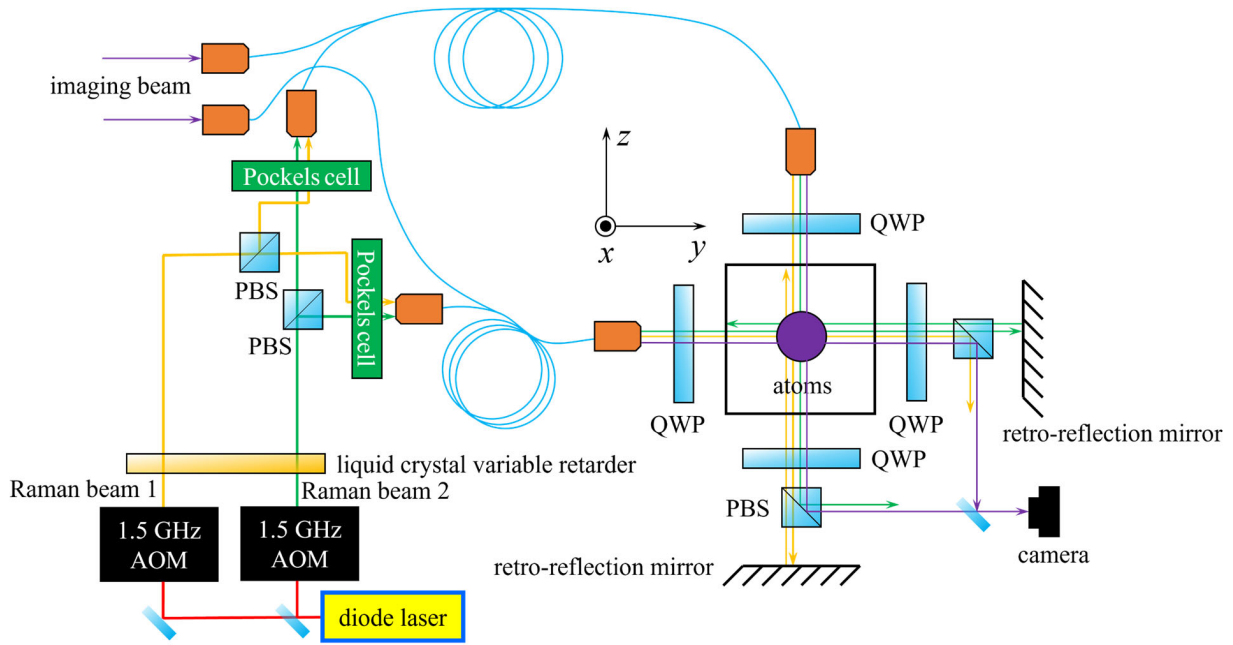

Figure 4. Schematic illustration of the PSI apparatus that can apply the Raman beams in two perpendicular directions. If the PSI employs $^{85}$Rb atoms, the energy difference between the two hyperfine ground states would be ~3.0 GHz. Therefore, the two Raman beams can be generated with two 1.5 GHz AOMs. Switching the direction of the Raman beams can be achieved by rotating the polarizations of the two Raman beams by 90° with a liquid crystal variable retarder. A Raman beam that originally passes the PBS will be reflected by the PBS if its polarization is rotated by 90°, and thus be directed to a different fiber port. Large momentum transfer (LMT) is realized by exchanging the polarizations of the two Raman beams with Pockels cells. The polarization of a Raman beam determines whether it will be retro reflected or directed away. By exchanging the polarizations of the two Raman beams, the one that is originally retro reflected, for example, will be directed away. In this way, the directions of the two Raman beams are reversed.

It should also be possible to realize a PSI apparatus where the Raman beams are applied in two orthogonal directions, with a separate retro-reflection mirror and a single detection system for both directions. By operating these two sets of Raman excitations alternatingly, it would be possible to measure rotation and acceleration along all three directions, thereby realizing an atom-interferometric inertial measurement unit (AIMU). This configuration is illustrated schematically in Figure 4. For concreteness, we assume here that the PSI employs $^{85}$Rb atoms, for which the energy difference between the two hyperfine ground states is ~3.0 GHz. Therefore, the two Raman beams can be generated with two 1.5 GHz AOMs. Switching the direction of the Raman beams can be achieved, for example, by rotating the polarizations of the two Raman beams by 90° with a liquid crystal variable retarder. A Raman beam that originally passes the PBS will be reflected by the PBS if its polarization is rotated by 90°, and thus be directed to a different fiber port.

Finally, we have noted that it should be possible to enhance significantly the sensitivity of both rotation sensing and accelerometry using this approach by making use of the technique of large momentum transfer.

## Acknowledgement:

This work has been supported equally in parts by NASA grant number 80NSSC20C0161, the Department of Defense Center of Excellence in Advanced Quantum Sensing under Army Research Office grant number W911NF202076, ONR grant number N00014-19-1-2181, and the U.S. Department of Energy, Office of Science, National Quantum Information Science Research Centers, Superconducting Quantum Materials and Systems Center (SQMS) under contract number DE-AC02-07CH11359.

# Appendix: Measurement of Acceleration Using the Three-Pulse PSI

Two problems are involved in measuring the acceleration parallel to the Raman beams with the three-pulse PSI: there will be no side peaks in the Fourier transform in the absence of the rotation and the phase of the Fourier transform depends on the choice of the origin of the coordinates. The solutions to these problems are as follows. In the absence of the rotation, the PSI would behave effectively as an ordinary light pulse interferometer, with the amplitude of the central peak corresponding to the total population of the atoms in state $|\uparrow\rangle$. The Raman phase of the last pulse in the PSI (i.e., the difference in the phase of the beat signal between the two legs of the Raman excitation fields in the last pulse with respect to the same for the first pulse), will then be scanned, while repeating the experiment many times. The displacement of the peak of the resulting fringe away from the zero Raman phase will then reveal the phase shift induced by the acceleration along the z-axis, which in turn will be used to determine the sign and amplitude of this acceleration, according to the relation $\phi_a = k_{\text{eff}} a_z T^2$. The second problem can be solved by repeating the experiment three times: once without a bias rotation, then with a bias rotation in the *x*-direction, and then with a bias rotation in the *y*-direction. The amplitudes and phases of the Fourier transforms of the signals resulting from these three cases can be used to determine unambiguously both components of rotation and the acceleration along the direction of the laser pulses, given the known amplitude and direction of the applied bias rotation. For simplicity, we again consider first the case where there is only an environmental rotation around the *x*-axis. With an arbitrary origin of the coordinates for the Fourier transform, the signal of the PSI becomes $\left(2 + e^{i[k_\Omega(y-y_0)+\phi_a]} + e^{-i[k_\Omega(y-y_0)+\phi_a]}\right)/4$ where $y_0$ is the coordinate of the initial position of the atoms. The positions of the peaks in the amplitude spectrum of the Fourier transform does not depend on the choice of the origin while the phase at the peak $k_y = k_\Omega$ is shifted by $(-k_\Omega y_0)$, meaning that only the measurement of the acceleration will be affected by the choice of the origin. Without the bias rotation, the peaks of the amplitude spectrum are located at $k_y = 0, \pm k_{\text{eff}}\Omega_{xe}T$, where $\Omega_{xe}$ is the environmental angular velocity in the x-direction. The phase at the peak $k_y = k_{\text{eff}}\Omega_{xe}T$ is $\phi_0 = (\phi_a - k_{\text{eff}}\Omega_{xe}Ty_0)$. With the bias rotation, the peaks are located at $k_y = 0, \pm k_{\text{eff}}(\Omega_{xe} + \Omega_{xb})T$, where $\Omega_{xb}$ is the bias angular velocity in the x-direction. The phase at the peak $k_y = k_{\text{eff}}(\Omega_{xe} + \Omega_{xb})T$ is $\phi_y = [\phi_a - k_{\text{eff}}(\Omega_{xe} + \Omega_{xb})Ty_0]$. It then follows that the acceleration induced phase shift can be found using the following equation:

$$\phi_a = \phi_0 - \frac{\Omega_{xe}}{\Omega_{xb}}(\phi_y - \phi_0) \tag{2}$$

In the presence of the environmental rotation along both the axes, a generalization of the derivation above shows that acceleration induced phase shift is given by:

$$\phi_a = \phi_0 - \frac{\Omega_{xe}}{\Omega_{xb}}(\phi_y - \phi_0) - \frac{\Omega_{ye}}{\Omega_{yb}}(\phi_x - \phi_0) \tag{3}$$

The overall protocol for determining rates of rotation around the x- and y-axes, and the acceleration along the z-axis, would work as follows. First, the experiment will be carried out without applying any bias rotation, and the two-dimensional Fourier transform of the detected signal would be carried out. If no sideband peaks are discernible, the technique of scanning the Raman phase of the last pulse will be adopted to measure $a_z$. If the sideband peaks are discernible after the operation of the PSI without any bias rotation, these would be used to infer the values of the rotations around the x- and y-axes: $\Omega_{xe}$ and $\Omega_{ye}$. Then the PSI would be operated again with a bias rotation around both the x- and y-axes, and Eq. (3) will be used to determine $a_z$.